\documentclass{ws-ijmpc_rev}
\usepackage{graphicx}
\begin{document}

\markboth{Hiroshi Koibuchi and Andrey Shobukhov}
{Crumpled-to-tubule transition of self-avoiding spherical meshwork}

\catchline{}{}{}{}{}

\title{Crumpled-to-tubule transition and shape transformations of a model of self-avoiding spherical meshwork
}

\author{Hiroshi Koibuchi
}

\address{Department of Mechanical and Systems Engineering, Ibaraki National College of Technology, Nakane 866 Hitachinaka, Ibaraki 312-8508, Japan
\\koibuchi@mech.ibaraki-ct.ac.jp
}

\author{Andrey Shobukhov}

\address{Faculty of Computational Mathematics and Cybernetics, Lomonosov Moscow State University, 
119991, Moscow, Leninskiye Gory, MSU, 2-nd Educaional Building, Russia
}

\maketitle

\begin{history}
\end{history}

\begin{abstract}
This paper analyzes a new self-avoiding (SA) meshwork model using the canonical Monte Carlo simulation technique on lattices that consist of connection-fixed triangles. The Hamiltonian of this model includes a self-avoiding potential and a pressure term.  The model identifies a crumpled-to-tubule (CT) transition between the crumpled and tubular phases. This is a second-order transition, which occurs when the pressure difference between the inner and outer sides of the surface is close to zero.  We obtain the Flory swelling exponents $\nu_{{\rm R}^2}(=\!D_f/2)$ and $\bar{\nu}_{\rm v}$ corresponding to the mean square radius of gyration $R_g^2$ and enclosed volume $V$, where $D_f$ is the fractal dimension. The analysis shows that $\bar{\nu}_{\rm v}$ at the transition is almost identical to the one of the smooth phase of previously reported SA model which has no crumpled phase.  

\end{abstract}

\ccode{PACS Nos.: 11.25.-w,  64.60.-i, 68.60.-p, 87.10.-e, 87.15.ak}
\section{Introduction}\label{Introduction}

A membrane can be regarded as a two-dimensional surface. Hence its mechanical strength is understood on notions based on the two-dimensional differential geometry \cite{HELFRICH-1973,POLYAKOV-NPB1986,NELSON-SMMS2004,WIESE-PTCP19-2000,FDAVID-SMMS2004}. The surface model of Helfrich and Polyakov has two different rotationally symmetric states: the smooth and crumpled phases. The smooth (crumpled) phase is expected in the model at the high (low) bending region $\kappa\!\to\!\infty$ ($\kappa\!\to\! 0$), where $\kappa[kT]$ is the bending rigidity.  The so-called crumpling transition between these two phases has been studied numerically \cite{KANTOR-KARDAR-NELSON-PRL1986,KANTOR-KARDAR-NELSON-PRA1987,KANTOR-NELSON-PRA1987,AMBJORN-NPB1993,Munkel-Heerman-PRL1995,KD-PRE2002} and theoretically \cite{Peliti-Leibler-PRL1985,PKN-PRL1988,DavidGuitter-EPL1988,NISHIYAMA-PRE-2004} for a long period of time. 

In contrast to the flat-to-crumpled transition, less is known about the crumpled-to-tubule (CT) transition. The tubular phase is characterized by an oblong surface shape, and hence the rotational symmetry is partly broken at the CT transition. Previous studies have reported the CT transition in phantom surfaces, which are surfaces with the ability to self-intersection \cite{Radzihovsky-Toner-PRL1995,Radzihovsky-Toner-PRE1998}. In those studies, an anisotropic bending rigidity is assumed in the local and internal directions of the surface. The CT transition has also been studied  using the non-perturbative renormalization group  formalization on phantom surfaces \cite{Kownacki-Mouhanna-2009PRE,Essafi-Kownack-Mouhanna-2011PRL}.  Moreover, the existence of the tubular phase was numerically shown in Ref. \refcite{BOWICK-etal-PRL1997}. The CT transition is of second-order on a phantom surface. Theoretical studies considered a self-avoiding (SA) interaction and identified the scaling relations for the tubule thickness and some other objects at the CT transition point \cite{Radzihovsky-SMMS2004}.  In addition, the experimental study with a partially polymerized membrane detected the transition to a wrinkling phase. The wrinkling phase found in this study is similar yet different from the tubular phase  \cite{Chaieb-etal-2006PRL,Chaieb-Malkova-Lal-2008JTB}.  The fractal dimension$D_f$ was measured at the wrinkling transition. Depending on the degrees of polymerization, $D_f$ is shown to have a value in the range $2.1\!\leq\!D_f\!\leq\!2.6$ at the wrinkling transition \cite{Chaieb-etal-2006PRL,Chaieb-Malkova-Lal-2008JTB}.  Hence, various theoretical, numerical and experimental studies support the presence of the CT transition. Yet, no studies have provided numerical changes associated with the CT transition on SA surfaces. 

The main problem associated with the CT transition on SA surfaces is that a SA surface model has no collapsed phase. This implies that the SA surface has no CT transition. For instance, the previously used model assumes a sheet with free boundaries without pressure term in the Hamiltonian \cite{BOWICK-TRAVESSET-EPJE2001}. 
In this paper we study whether a SA surface model with sphere topology undergoes a CT transition. The introduced SA property defines well the volume enclosed by the surface, and hence, the pressure difference between the inside and outside of the surface is controlled. Moreover, the model spontaneously generates a tubular phase by breaking the rotationally symmetrical structure. The model exhibits a marked change in the Flory swelling coefficients at the CT transition, which separates the wrinkled phase and the tubular phase at small bending region. This wrinkled phase is characterized by $D_f\simeq 2.2$. The wrinkled phase is considered to be almost smooth and it seems to correspond to the phase at $\kappa\!\to\! 0$ in the SA model of Ref. \refcite{BOWICK-TRAVESSET-EPJE2001}. 
  
\section{Model}\label{model}

\subsection {Continuous Model} 
We start with the continuous model. In the string model context, a membrane is represented by a mapping $X:M\ni x\mapsto {\bf r}(x)\in {\bf R}^3$, where $M$ is a two-dimensional surface of sphere topology and $x\!=\!(x_1,x_2)$. The image $X(M)(\subset {\bf R}^3)$ corresponds to a membrane. Using this symbol $X$, the continuous partition function $Z_c$ is written as
\begin{eqnarray} 
\label{Part-Func-cont}
Z_c = \int {\mathcal D}X \exp\left[-\frac{1}{k_BT}S_c({\bf r})\right], \quad S_c({\bf r})=\gamma S^c_1 + \kappa S^c_2 -{\it \Delta}p \,V+ bU_c,
\end{eqnarray}
where  $S_c({\bf r})$ is the continuous Hamiltonian. The parameters $\gamma$, $\kappa$ and $b$ in $S_c$ denote the surface tension coefficient, the bending rigidity, and the excluded volume parameter, respectively. 
Here ${\it \Delta}p$ is the pressure difference between the inside and outside of the surface defined by ${\it \Delta}p\!=\!p_{\rm in}\!-\!p_{\rm out}$. A positive (negative) ${\it \Delta}p$ implies that the inside pressure is greater (smaller) than the outside pressure. The volume $V$ is set positive for the self-avoiding surfaces, while for the phantom surfaces $V$ can be negative. 

The energies $S^c_1$, $S^c_2$, and $U_c$ are given by
\begin{eqnarray} 
\label{Cont-energies}
&&S^c_1({\bf r})=\int\sqrt{g}d^2x, \quad S^c_2=\frac{1}{2}\int\sqrt{g} d^2x \left(g^{ab}\frac{\partial {\bf t}_a}{\partial x_b}\right)^2, \nonumber \\
&&U_c=\frac{1}{2}\int d^2x\int d^2x^\prime \delta\left({\bf r}(x)-{\bf r}(x^\prime)\right),
\end{eqnarray}
where $g$ is the determinant of the metric $g_{ab}$, and $g^{ab}$ is its inverse, and  ${\bf t}_a (=\!(\partial {\bf r}/\partial x_a)/\sqrt{\partial {\bf r}/\partial x_a})$ in $S_2^c$ denotes a unit tangential vector of the membrane. $S_1^c$ is the area of the membrane, and we call $S_1^c$ the area energy. We assume the Euclidean metric 
\begin{eqnarray}
g_{ab}=\delta_{ab}\quad {\rm in} \; S_2^c, 
\end{eqnarray}
then we have
\begin{eqnarray}
\label{Eucl-S2}
S_2^c  
&&=\frac{1}{2}\int d^2x\left[\left({\partial_1 {\bf t}_1} \right)+\left({\partial_2 {\bf t}_2} \right)\right]^2 \nonumber \\
&&=\frac{1}{2}\int d^2x\left[\left({\partial_1 {\bf t}_1} \right)^2\!+\!\left({\partial_2 {\bf t}_2} \right)^2\!+\!2\left({\partial_1 {\bf t}_1} \right)\cdot\left({\partial_2 {\bf t}_2} \right)\right].
\end{eqnarray}
Note that replacing ${\bf t}_a$ by $\partial {\bf r}/\partial x_a$ in $S_2^c$ we get $(1/2)\int (\partial^2 {\bf r})^2$.  This term $(1/2)\int (\partial^2 {\bf r})^2$ is like the one in the curvature energy of the Ginzburg-Landau Hamiltonian for membranes \cite{FDAVID-SMMS2004}. For this reason we shall simply call $S_2^c$ the curvature energy. The final term $U_c$ in Eq. (\ref{Cont-energies}) represents a self-avoiding interaction between two points of the membrane and is an extension of Doi-Edwards model for polymer \cite{Doi-Edwards-1986}.

\subsection {Discrete Model} 
The discrete model is obtained from the continuous model introduced in the previous subsection and is defined on a triangulated sphere, which is obtained by splitting the icosahedron \cite{KOIB-EPJB-2007-3}. The coordination number $q$ of vertices is $q\!=\!6$ at almost all vertices except $q\!=\!5$ at 12 vertices. 

The discrete partition function of the model is given by
\begin{equation} 
\label{Part-Func}
 Z = \int^\prime \prod _{i=1}^{N} d {\bf r}_i \exp\left[-S({\bf r})\right], 
\end{equation}
where the prime in $\int^\prime \prod _{i=1}^{N}d {\bf r}_i$ denotes that the three-dimensional multiple integrations are performed by fixing the center of mass of the surface to the origin of ${\bf R}^3$. The parameter $k_BT$ in the Boltzmann factor is fixed to $k_BT\!=\!1$ for simplicity.  The Hamiltonian $S({\bf r})$ looks as follows:
\begin{eqnarray}
\label{Disc-Eneg} 
&& S({\bf r})=S_1 +  \kappa S_2 -{\it \Delta}p \,V+ U, \quad S_1=\sum_{\it \Delta} A_{\it \Delta},\nonumber \\
&&S_2=\frac{1}{3}\sum_{ij}\left({\bf t}_i- {\bf t}_j\right)^2+\frac{1}{3}\sum_{(ij),(kl)}\left({\bf t}_i- {\bf t}_j\right)\cdot\left({\bf t}_k- {\bf t}_l\right),  \nonumber\\
&& U=\sum_{{\it \Delta},{\it \Delta}^\prime} U({\it \Delta},{\it \Delta}^\prime),  \quad
U({\it \Delta},{\it \Delta}^\prime)= \left\{
    \begin{array}{@{\,}ll}
    \infty & \; ({\rm triangles}\; {\it \Delta},{\it \Delta}^\prime\;{\rm intersect}) \\
         0 & \; ({\rm otherwise}). 
    \end{array}
    \right. 
\end{eqnarray}

The surface tension coefficient $\gamma$ can always be fixed to $\gamma\!=\!1$ in $S\!=\!\gamma S_1\!+\!\kappa S_2\!+\!{\it \Delta}p \,V\!+\! U$. Indeed, the scale invariance of $Z$ allows us to rescale the variable ${\bf r}\!\to\! {\bf r}^\prime\!=\!\sqrt{\gamma}{\bf r}$ in $Z$.  As $S_2$ and $U$ are scale independent, 
$$Z\!=\!\int^\prime \prod_{i=1}^{N}d{\bf r}_i \exp\left[ -\left(\gamma S_1({\bf r})\!+\!\kappa S_2-\!{\it \Delta}p \,V({\bf r})\!+\! U\right)\right]$$
 can also be written as
 $$Z=\int^\prime \prod_{i=1}^{N}d{\bf r}^\prime_i \exp\left[ -\left(S_1({\bf r}^\prime)\!+\!\kappa S_2-\!{\it \Delta}p^\prime \,V({\bf r}^\prime)\!+\! U\right)\right],$$
 which would be identical with the original $Z$  up to a multiplicative constant if we replace ${\it \Delta}p^\prime\!=\!\gamma^{-3/2}{\it \Delta}p$ by ${\it \Delta}p$. 
The excluded volume parameter $b$ is suppressed in $U$ of Eq. (\ref{Disc-Eneg}).

\begin{figure}[!h]
\centering
\includegraphics[width=9.5cm]{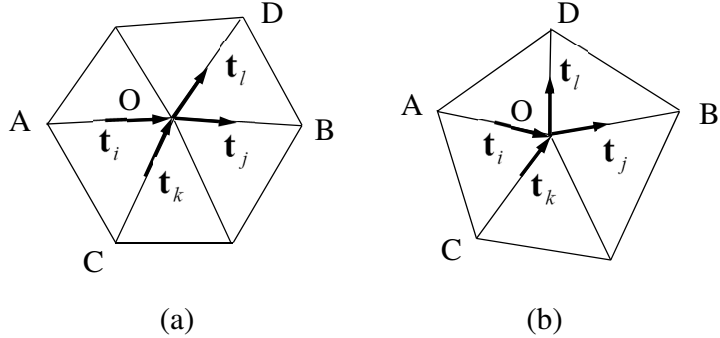}  
\caption{ A possible configuration of the product $({\bf t}_i\!-\!{\bf t}_j)\cdot({\bf t}_k\!-\!{\bf t}_l)$ for the second term of $S_2$ at (a) the $q\!=\!6$ vertex and (b) at the $q\!=\!5$ vertex.  We have three (or five) possible configurations for $({\bf t}_i\!-\!{\bf t}_j)\cdot({\bf t}_k\!-\!{\bf t}_l)$ in (a) (or in (b)). 
 } 
\label{fig-1}
\end{figure}

The vectors ${\bf t}_i$ and ${\bf t}_j$ in the first term of $S_2$ are those on a diagonal line of a hexagonal lattice (Fig. \ref{fig-1}(a)), where we have three possible pairs ${\bf t}_i\!-\!{\bf t}_j$ and include them in the sum of the first term. The pair ${\bf t}_j-{\bf t}_i$ corresponds to the partial derivative ${\partial {\bf t}_a}/{\partial x_a}$ in the continuous $S_2^c$ in Eq. (\ref{Eucl-S2}). In the second term of $S_2$, ${\bf t}_i\!-\!{\bf t}_j$ and ${\bf t}_k\!-\!{\bf t}_l$ are those shown in Fig. \ref{fig-1}(a), where the triangles $OAC$ and $OBD$ are opposite to each other. Three possible inner products $({\bf t}_i\!-\!{\bf t}_j)\cdot({\bf t}_k\!-\!{\bf t}_l)$ are included in the sum, because we have three different pairs of triangles like $OAC$ and $OBD$ on a hexagonal lattice. The factor $1/3$ is included in $S_2$ of Eq. (\ref{Disc-Eneg}) because every vertex is assumed to be the center of hexagon and therefore the summation is triply duplicated. On a pentagonal lattice such as shown Fig. \ref{fig-1}(b), we have five possibilities for ${\bf t}_i\!-\!{\bf t}_j$, which are included in the first term of $S_2$ by modifying the coefficient $1/3$ to $1/6$. We also have five different products $({\bf t}_i\!-\!{\bf t}_j)\cdot({\bf t}_k\!-\!{\bf t}_l)$ for the second term of $S_2$ on a pentagonal lattice, and we include those in the second term of $S_2$ with the coefficient $1/6$. 

The area energy $S_1$ influences only the area constant and does not always suppress elongation of triangles. This situation is in striking contrast to a model based on the Gaussian bond potential $S_1\!=\!\sum_{ij}({\bf r}_i\!-\!{\bf r}_j)^2$. However, the curvature energy $S_2$ has a resistance against in-plane deformations of triangles at  all vertices  due to the second term of $S_2$. Moreover, the first term of $S_2$ prohibits the bonds $i$ and $j$ from in-plane bending. This in-plane bending resistance is seen along the diagonal axes (see Fig. \ref{fig-1}(a)). Therefore, the model in this paper is different from fluid surface models, where no in-plane bending resistance is seen, although elongated surfaces are expected to appear.

The sum $\sum_{{\it \Delta}{\it \Delta}^\prime}$ in the self-avoiding potential $U$ denotes the sum over all pairs of non-nearest neighbor (or disjointed) triangles ${\it \Delta}$ and ${\it \Delta}^\prime$. The potential $U({\it \Delta},{\it \Delta}^\prime)$ is defined in such a way that ${\it \Delta}$ and ${\it \Delta}^\prime$ do not intersect each other. The SA interaction defined by $U$ in Eq. (\ref{Disc-Eneg})  slightly differs from the one assumed in the SA model of Bowick et.al. \cite{BOWICK-TRAVESSET-EPJE2001}, where the triangles are allowed to self-intersect with small probability. The SA interaction in the model of Bowick et.al. \cite{BOWICK-TRAVESSET-EPJE2001} is more close to $U_c$ in Eq. (\ref{Cont-energies}) and is considered to be an impenetrable plaquette model. Although the model in this paper may also be regarded as an impenetrable plaquette model, the definition of $U$ in Eq. (\ref{Disc-Eneg}) is simpler than those in the model of Bowick et.al. \cite{BOWICK-TRAVESSET-EPJE2001}.  The definition of $U$ in  Eq. (\ref{Disc-Eneg}) also differs from the SA interaction of the beads-and-springs model \cite{KANTOR-KARDAR-NELSON-PRL1986,KANTOR-KARDAR-NELSON-PRA1987}. However, these differences should not influence the final outcomes, such as the swelling exponents. 

\section{Monte Carlo technique}\label{MC-Techniques}
\begin{figure}[!h]
\centering
\includegraphics[width=5.0cm]{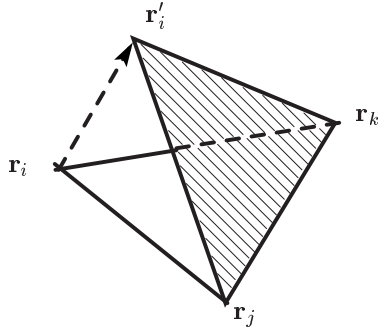}  
\caption{The SA interaction prohibits (i) the shaded triangle from intersecting with the bonds which are disjointed to the triangle, and it also prohibits (ii) the bonds ${\bf r}_j-{\bf r}_i^\prime$ and ${\bf r}_k-{\bf r}_i^\prime$  from intersecting with the triangles which are disjointed to the bonds. 
 } 
\label{fig-2}
\end{figure}
The canonical Metropolis Monte Carlo (MC) technique is used to update the variable ${\bf r}$.  
The constraint $U({\it \Delta},{\it \Delta}^\prime)$ in Eq. (\ref{Disc-Eneg}) is imposed on the triangles ${\it \Delta}$ and ${\it \Delta}^\prime$ as follows: $i$, $j$ and $k$ represent the vertices of a triangle, while ${\bf r}_i$ and ${\bf r}_i^\prime$ denote the current and new positions of the vertex $i$ (Fig. \ref{fig-2}). As the vertex $i$ moves from ${\bf r}_i$ to ${\bf r}_i^\prime$, a new triangle emerges (Fig. \ref{fig-2}). The self-avoiding interaction is implemented by testing whether the shaded triangle intersects with all other bonds. Since all bonds are edges of triangles, the self-avoidance between bonds and triangles automatically prohibits the intersections of bonds with bonds. The violations in this rule would disjoint the connecting triangles. All the neighboring triangles that share the vertex $i$ should be taken into account simultaneously. The other task for the implementation is to test whether the bonds ${\bf r}_j-{\bf r}_i^\prime$ and ${\bf r}_k-{\bf r}_i^\prime$ intersect with all other triangles. 

We assume a sphere of radius $R_0$ at the center of mass of the triangle and test for the SA properties within the sphere (Fig. \ref{fig-2}).  The radius $R_0$ is assumed to be $R_0\!=\!L_{\rm max}$, where $L_{\rm max}$ is the maximum bond length computed every 1 MC sweep (MCS). We also check whether the disjointed triangles intersect with each other at every 500 MCSs. No intersection is found at any bending rigidity even under a negative pressure such as ${\it \Delta}p\!=\!-0.5$.  

The total number of MCS after the thermalization is about $2\times 10^7\sim 3\times 10^7$ for the surface with  $N\!=\!1962$. A relatively small number of MCS is assumed on smaller surfaces. The total number of the thermalization MCS is about $0.5\times 10^6$. The thermalization MCS in the tubular phase is very large; it is sometimes $1\times 10^7$ or more at the phase boundary close to the planar phase on the $N\!=\!1442$ surface.  

\section{Results}\label{Results}
\subsection{Under the pressures ${\it \Delta}p\!=\!0$ and ${\it \Delta}p\!=\!-0.5$}\label{negative_p}

\begin{figure}[!h]
\centering
\includegraphics[width=10.5cm, clip]{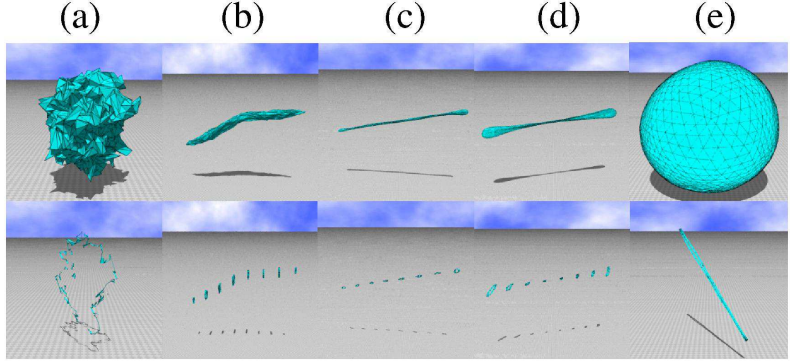}  
\caption{(Color on-line) The snapshots of surfaces and the surface sections of size $N\!=\!1442$ obtained under ${\it \Delta}p\!=\!0$ at (a) $b\!=\!0.1$ (collapsed), (b) $b\!=\!1$ (tubular), (c) $b\!=\!5$ (tubular), (d) $b\!=\!50$ (tubular), (e) $b\!=\!100$ (planar).} 
\label{fig-3}
\end{figure}
Figures \ref{fig-3}(a)--\ref{fig-3}(e) illustrate snapshots of surfaces and surface sections under the zero pressure condition ${\it \Delta}p\!=\!0$. The surface size is $N\!=\!1442$. The assumed bending rigidities are in the range $0.1\!\leq\! \kappa\!\leq\! 100$. The scales of the figures are all different from each other. The snapshots in Fig. \ref{fig-3}(a) indicate that the surface is not highly crumpled, however the surface becomes more crumpled at $\kappa\to 0$ under ${\it \Delta}p\!=\!-0.5$. The phase structure at ${\it \Delta}p\!=\!-0.5$ is almost identical to the one at ${\it \Delta}p\!=\!0$ except for the crumpled phase. 
 We should note that the collapsed surface disappears even at $\kappa\to 0$ when ${\it \Delta}p\!\simeq\!0$. This is consistent with the previous result that the SA sheet has no crumpled phase \cite{BOWICK-TRAVESSET-EPJE2001}.   

\begin{figure}[t]
\centering
\includegraphics[width=10.5cm]{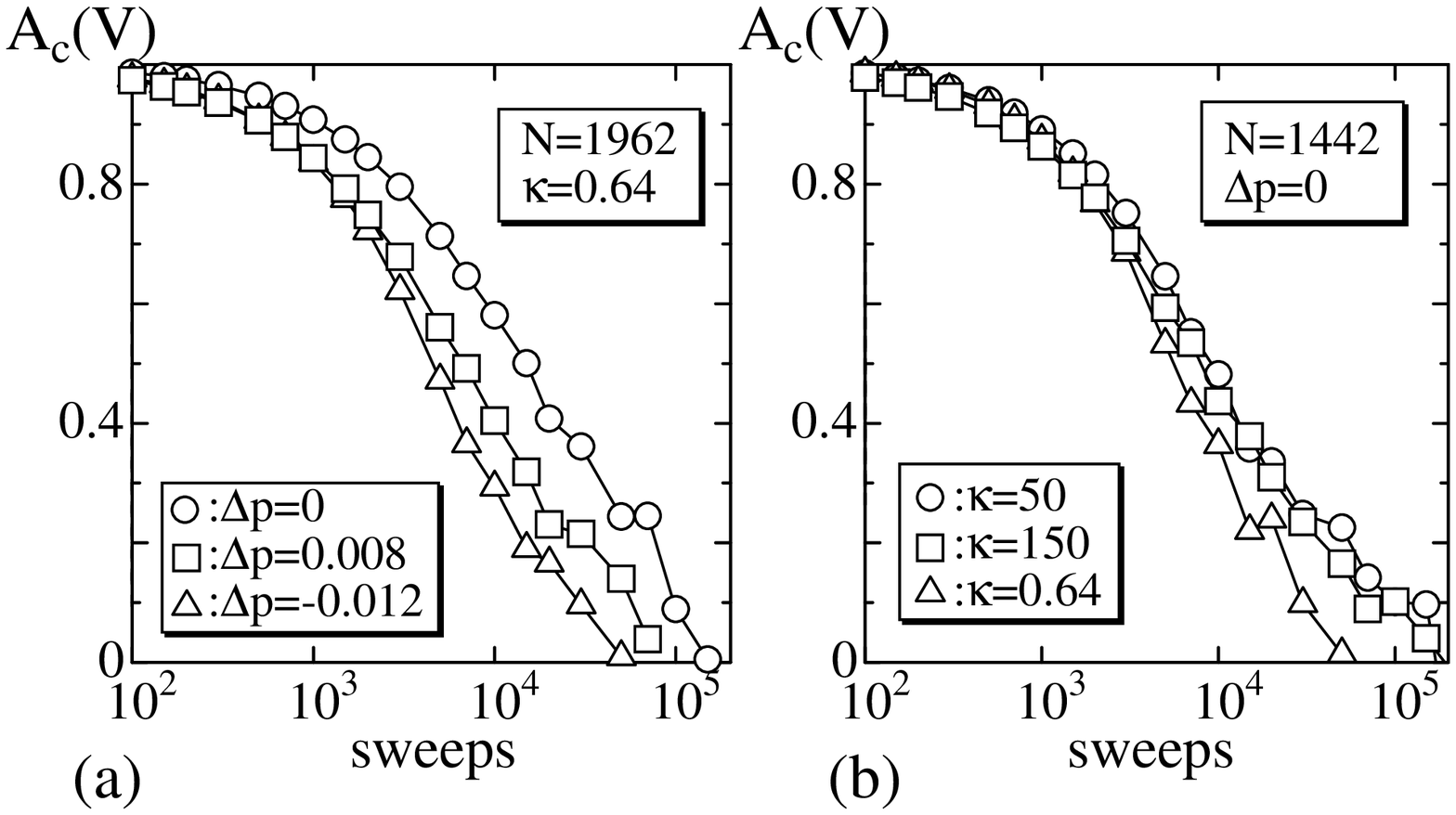}  
\caption{The autocorrelation coefficient $A_c(V)$ of the enclosed volume vs. MCS of (a) the $N\!=\!1962$ surface at the CT transition point and (b) the $N\!=\!1442$ surface at ${\it \Delta} p\!=\!0$.  } 
\label{fig-4}
\end{figure}
The decorrelation time for the enclosed volume $V$ can be estimated with the help of the autocorrelation coefficient defined as
\begin{eqnarray}
\label{Acor} 
 A_c(V)(n)=\frac{{\sum_i V(i)V(i+n)}}{\sqrt{\sum_i \left[V(i)\right]^2}\sqrt{\sum_i \left[V(i+n)\right]^2}},\quad (n=1,2,\cdots),
\end{eqnarray}
where $\{V(i)\}$ denotes a series of data obtained every $50$ MCS after the thermalization MCS ($n\!=$MCS/50).  We see that $A_c(V)\!\simeq\! 0$ at $1\times 10^5$ MCS at the CT transition on the $N\!=\!1962$ surface (Fig. \ref{fig-4}(a)). This implies that the total number of MCS ($2\times 10^7\sim 3\times 10^7$) is sufficient for measurements. It is also seen on the $N\!=\!1442$ surface that $A_c(V)\!\simeq\! 0$ at the same order of MCS ($2\times 10^5$) in the region $\kappa \!\leq\! 150$ at least. The decorrelation time at the tubular phase ($\kappa\!=\!50$) is considered to be quite small in comparison with the thermalization MCS ($\sim 1\!\times\! 10^7$), which is necessary only for the shape change from the initial configuration (sphere) to tubular (or planar) surface shown in Fig. \ref{fig-3}(d) (or \ref{fig-3}(e)).

We should emphasize that there is no sphere phase under  ${\it \Delta}p\!=\!0$ and ${\it \Delta}p\!=\!-0.5$ at least. This is true even in the model without the SA interaction. 
\begin{figure}[t]
\centering
\includegraphics[width=10.5cm]{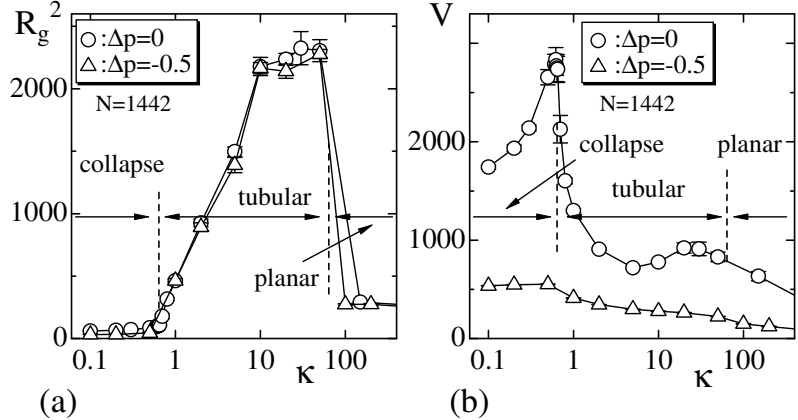}  
\caption{(a) The mean square radius of gyration $R_g^2$  vs. $\kappa$ and (b) the enclosed volume $V$ vs. $\kappa$. Vertical dashed lines represent the phase boundaries. The solid lines connecting the data symbols are drawn to guide the eyes.} 
\label{fig-5}
\end{figure}
The mean square radius of gyration {$R_g^2$}  is defined by
\begin{equation}
\label{X2}
R_g^2=\frac{1}{N} \sum_i \left({\bf r}_i-\bar {\bf r}\right)^2, \quad \bar {\bf r}=\frac{1}{N} \sum_i {\bf r}_i,
\end{equation}
where $\bar {\bf r}$ is the center of mass of the surface. The value of $R_g^2$ changes depending on the distribution of the vertices in ${\bf R}^3$, and hence $R_g^2$  as well as the enclosed volume $V$ can reflect the shape transformations. However, the quantities $R_g^2$  and $V$ show two different behaviors against $\kappa$. Figure \ref{fig-5}(a) shows $R_g^2$  vs. $\kappa$ under ${\it \Delta}p\!=\!0$ and ${\it \Delta}p\!=\!-0.5$. We observe that $R_g^2$  discontinuously changes at the phase boundaries between the planar and tubular phases. The change in $R_g^2$ reflects transitions from the tubular phase to either the collapsed or planar states. In contrast, the alteration in $V$ reveals a transition between the tubular and collapsed phases only under ${\it \Delta}p\!=\!0$. The reason why $V$ has a peak at the boundary between the collapsed and tubular phases under ${\it \Delta}p\!=\!0$ is that the surface is relatively inflated at the transition point as we will see below.

\begin{figure}[!t]
\centering
\includegraphics[width=10.5cm]{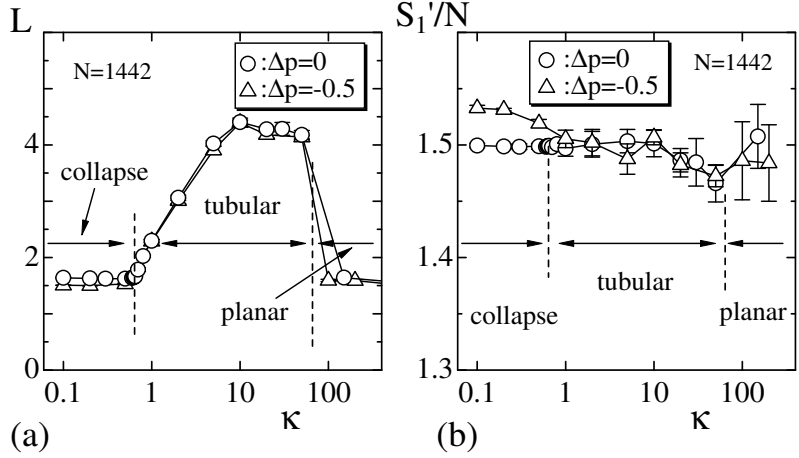}  
\caption{(a) The mean bond length $L$ vs. $\kappa$, and (b) $S_1^\prime/N$ vs. $\kappa$ under ${\it \Delta}p\!=\!0$ and ${\it \Delta}p\!=\!-0.5$, where $S_1^\prime\!=\!S_1\!-\!(3/2){\it \Delta}p\,V$. } 
\label{fig-6}
\end{figure}

The phase transition is also associated with the structure of triangles:  the surface consists of equilateral triangles in the smooth spherical phase, whereas it includes extremely-oblong triangles in the tubular phase. Figure \ref{fig-6}(a) shows the mean bond length  $L$ vs. $\kappa$. The discontinuous change in $L$ at the phase boundaries clearly indicates that the phase transitions are accompanied by a structural change of surfaces. This structural change causes the apparent separation of the planar phase from the tubular phase by a first-order transition. 

It is nontrivial that the bond lengths remain finite, because both $S_1$ and $S_2$ are defined independently of the bond length. In fact, the bond length becomes infinitely long in the model of which Hamiltonian is given by $S\!=\!S_1\!+\!\kappa S_3$, where $S_1$ is the area energy in Eq. (\ref{Disc-Eneg}) and $S_3\!=\!\sum_{ij}(1\!-\!{\bf n}_i\cdot{\bf n}_j)$ \cite{ADF-NPB1985}. One can easily check that this model is numerically ill-defined. In this ill-defined model, both $S_1$ and $S_3$ are independent of the bond length just like in the model of this paper. As mentioned previously, the curvature energy $S_2$ in this paper resists an in-plane bending while the bending energy $S_3$ does not. Therefore, an in-plane bending energy component included in $S_2$ is expected to make the area energy model well-defined.  

Figure \ref{fig-6}(b) shows $[S_1\!-\!(3/2){\it \Delta}p\,V]/N$, denoted by $S_1^\prime/N$, vs. $\kappa$. Because of the scale invariance of $Z$ in Eq. (\ref{Part-Func}), $S_1^\prime/N$ would be $S_1^\prime/N\!=\!3/2$ for sufficiently large $N$.  The scale invariance of $Z$ is represented by $\partial_\alpha Z(\alpha {\bf r})/\partial \alpha |_{\alpha=1} \!=\!0$, where $\alpha$ is a multiplicative factor of ${\bf r}$ as a scale transformation \cite{WHEATER-JP1994}. As we have seen in Section \ref{model}, this transformation changes $S_1$ and $V$ to $\alpha^2 S_1$ and $\alpha^3 V$, while $S_2$ and $U$ remain unchanged.  Furthermore, the integration $\int \prod_i d {\bf r}_i$ also changes to $\alpha^{3(N\!-\!1)}\int \prod_i d {\bf r}_i$.  Thus, the relation $S_1^\prime/N\!=\!3/2$ is achieved in the limit of $N\!\to\!\infty$. The results obtained in the range $0.1\!\leq\! \kappa\!\leq\! 200$ are consistent with this prediction with minor exceptions.  At the edge of the tubular phase towards the planar phase, we see a small deviation of $S_1^\prime/N$ from $3/2$. This deviation comes from the fact that the vertices distribute almost one-dimensionally on the tubular surfaces, where the transformation of $V$ is not always according to the rule $V\!\to\!\alpha^3 V$. A deviation of $S_1^\prime/N$ from $3/2$ is also seen at small bending region when ${\it \Delta}p\!=\!-0.5$. This also comes from the fact that the movement of vertices is constrained because the SA surface is collapsed, and the transformation $V\!\to\!\alpha^3 V$ is expected to be slightly broken.    

\begin{figure}[!t]
\centering
\includegraphics[width=10.5cm]{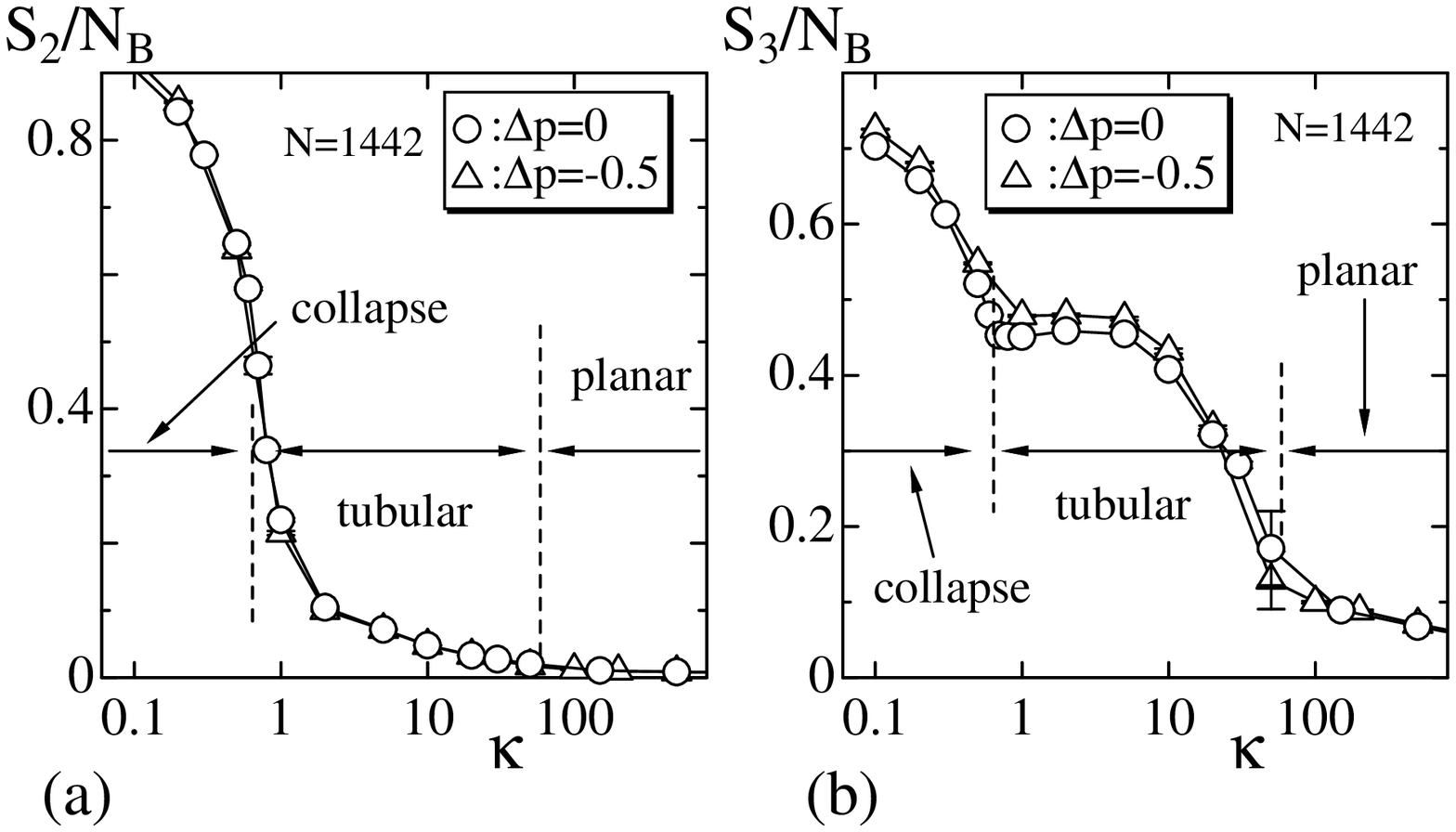}  
\caption{(a) The curvature energy $S_2/N_B$ vs. $\kappa$, and (b) $S_3/N_B$ vs. $\kappa$ under ${\it \Delta}p\!=\!0$ and ${\it \Delta}p\!=\!-0.5$. where $N_B$ is the total number of bonds, and $S_3\!=\!\sum_{ij}(1\!-\!{\bf n}_i\cdot{\bf n}_j)$.} 
\label{fig-7}
\end{figure}
We expect that $S_2/N_B$ discontinuously changes between the tubular and planar phases although the discontinuity ${\it \Delta} S_2/N_B$ is very small (Fig.{\ref{fig-7}}(a)). The discontinuity is hardly seen in the plot.  The reason why ${\it \Delta} S_2/N_B$ is very small is that the bending rigidity $\kappa$ at the transition is very large.  On the other hand, the bending energy $S_3\!=\!\sum_{ij}(1\!-\!{\bf n}_i\cdot{\bf n}_j)$, which is not included in the Hamiltonian, rapidly changes at the phase boundary (Fig.{\ref{fig-7}}(b)). This implies that the surface smoothness rapidly changes at the phase boundary. Note that the total number of bond pairs at which $S_2$ is defined is identical to $N_B\!=\!3N\!-\!6$; the total number of bond pairs is $3 (5/2)$ for the $q\!=\!6 (q\!=\!5)$ vertices.

\subsection{Under the small bending rigidity $\kappa\!=\!0.64$}\label{positive_b}
In the previous subsection, we saw clear separations of the states: planar and tubular phases.  However,  the conditions under which the transition between the crumpled and tubular phases occurs remains unclear. In this subsection, we firstly clarify the order of the CT transition by varying the pressure difference ${\it \Delta}p$ while fixing $\kappa$ at $0.64$. The reason why $\kappa$ is fixed to $\kappa\!=\!0.64$ is because the variance 
\begin{equation}
C_V=\frac{1}{N^{3/2}}\left< \left(V\!-\!\langle V\rangle\right)^2\right>
\end{equation}
has a peak at $\kappa\!\simeq\!0.64$ under ${\it \Delta}p\!=\!0$. Note that we use $N^{3/2}$ in place of $N$ in the definition of $C_V$. This is because the enclosed volume $V$ is proportional to $N^{3/2}$ if the surface is smooth and spherical.  

\begin{figure}[!h]
\centering
\includegraphics[width=9.5cm]{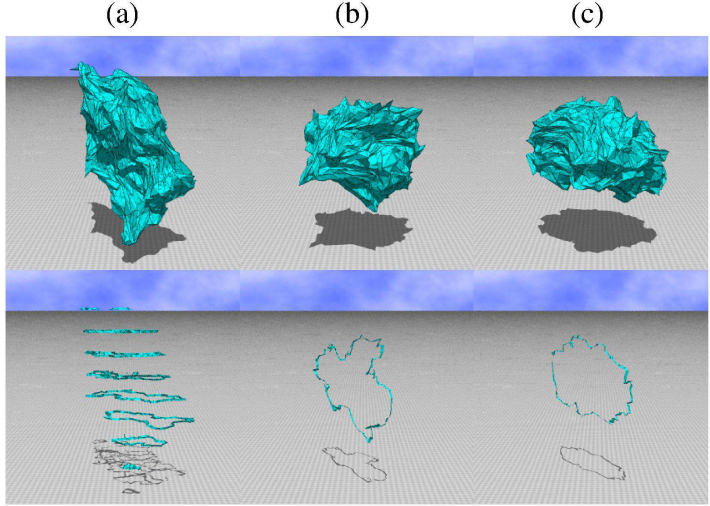}  
\caption{(Color on-line) The snapshots of surfaces and the surface sections of size $N\!=\!1962$ obtained at   (a)  ${\it \Delta}p\!=\!-0.004$ (tubular), (b)  ${\it \Delta}p\!=\!0.002$  (crumpled/tubular), and (c)  ${\it \Delta}p\!=\!0.012$ (crumpled)  under $\kappa\!=\!0.64$. The scales of the figures are the same.} 
\label{fig-8}
\end{figure}
Figure \ref{fig-8} shows the snapshots of surface and surface section obtained at ${\it \Delta}p\!=\!-0.004$, $0.002$, and $0.012$. Because of the SA potential, as we mentioned above, the surface does not completely collapse under ${\it \Delta}p\!\simeq\!0$. Nevertheless, we use the terminology {\it crumpled} for the surface state obtained at ${\it \Delta}p\!=\!0.012$ (Fig. \ref{fig-8}(a)). Indeed, the surface under this condition contains more wrinkles than the surface in the planar phase. Furthermore, the surface in the crumpled phase is symmetric under arbitrary 3-dimensional rotation.  In contrast, the surface in the tubular (or planar) phase is symmetric under the rotation only around an axis which is spontaneously generated.

\begin{figure}[!t]
\begin{center}
\includegraphics[width=10.5cm]{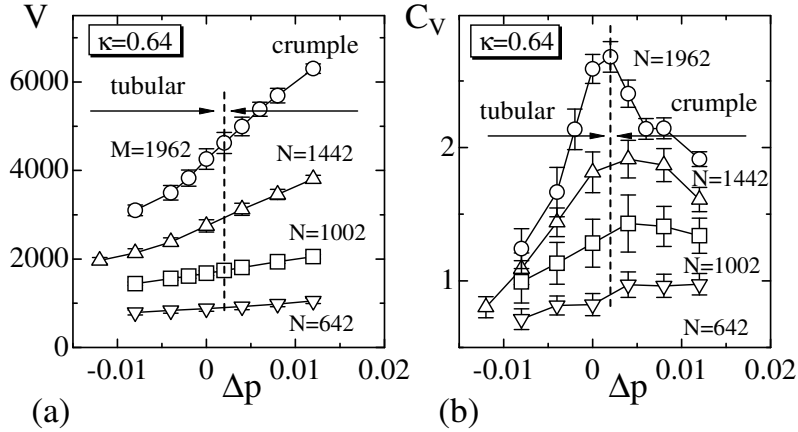}  
\caption{(a) The enclosed volume $V$ vs. ${\it \Delta}p$ and (b) $C_V$ vs. ${\it \Delta}p$ under $b\!=\!0.64$. The vertical dashed line denotes the phase boundary.} 
\label{fig-9}
\end{center}
\end{figure}
Figure \ref{fig-9}(a) shows $V$ vs. ${\it \Delta}p$ obtained on the surfaces with $N\!=\!642\sim N\!=\!1962$. The variance $C_V$ is plotted in Fig. \ref{fig-9}(b). We find that $C_V$ has a peak at ${\it \Delta}p\!\simeq\!0.002$, which is very close to ${\it \Delta}p\!=\!0$. This peak position represents the phase boundary between the crumpled and tubular phases. 

\begin{figure}[!b] 
\centering
\includegraphics[width=10.5cm]{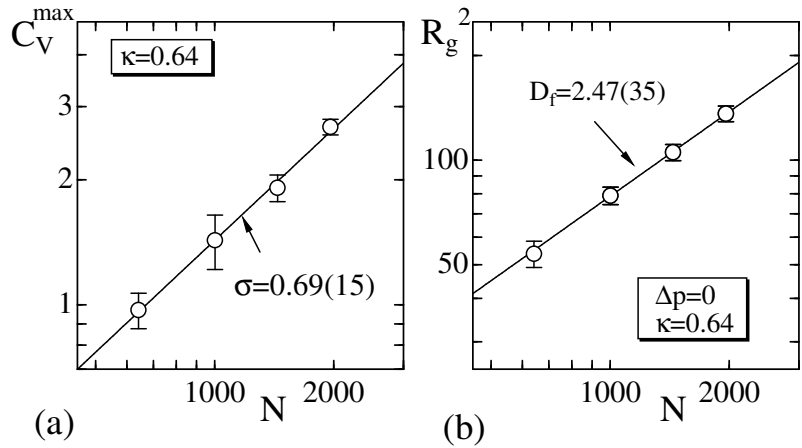}  
\caption{(a) $C_V^{\rm max}$ vs. $N$ obtained at several different values of ${\it \Delta}p$ close to  ${\it \Delta}p\!=\!0$, (b) $R_g^2$ vs. $N$ under ${\it \Delta}p\!=\!0$. Both figures are plotted in log-log scales.} 
\label{fig-10}
\end{figure}
The peak values $C_V^{\rm max}$ are plotted against $N$ in a log-log scale in Fig. \ref{fig-10}(a). The straight line is drawn by fitting the data to 
\begin{equation}
\label{CV_scale}
C_V^{\rm max}\sim N^{\frac{3}{2}\sigma}, \quad \sigma=0.69\pm0.15, \quad (N\to \infty).
\end{equation}
The slope of the line is given by $(3/2)\sigma$, where $\sigma$ a critical exponent. The result $\sigma\!=\!0.69(15)$ indicates that the CT transition is of second order. In order to compare our result with the first-order transition in the previous study \cite{Dammann-etal-JPIF1994}, we perform the fitting $\langle \left(V\!-\!\langle V\rangle\right)^2\rangle\sim N^\lambda$ and find $\lambda\!=\!2.42(9)$. The obtained value is significantly smaller than $\lambda\!=\!3.62(2)$ \cite{Dammann-etal-JPIF1994}. 

Figure \ref{fig-10}(b) shows $R_g^2$ vs. $N$ in a log-log scale. The data are obtained at ${\it \Delta}p\!=\!0$. The straight lines are drawn by fitting the data to  
\begin{equation}
\label{X2_scale}
R_g^2\sim N^{\nu_{{\rm R}^2}}= N^{\frac{2}{D_f}} \quad (N\to \infty),
\end{equation}
where $D_f$ is the fractal dimension of the surface.  The large three data sets are used in the fitting in Fig. \ref{fig-10}(b). The result $D_f\!=\!2.47(35)$ is interesting, because it is also close to $H\!=\!2.59(57)$ in the crumpled phase close to the crumpling transition of the canonical surface model, which is allowed to self-intersect \cite{KOIB-PRE-2005}.  This is the reason why we call this transition  the CT transition despite the surface in the crumpled phase is not always crumpled, at least under ${\it \Delta}p\!=\!0$.    
The results obtained in this paper including the swelling exponent $\nu_{{\rm R}^2}(=\!2/D_f)$ are shown in Table \ref{table-1}.
\begin{table}[hbt]
\tbl{Fractal dimension $D_f$, and the swelling exponents $\nu_{{\rm R}^2}$,  $\bar{\nu}_{\rm v}$, and $\nu_{\rm v}$,  obtained at ${\it \Delta}p\!=\!0.012\sim{\it \Delta}p\!=\!-0.008$ under $\kappa\!=\!0.64$. }
 {\begin{tabular}{ccccc}
\hline
   ${\it \Delta}p$ &  $D_f$ &  $\nu_{{\rm R}^2}$ &  $\bar{\nu}_{\rm v}$ &  $\nu_{\rm v}$ \\
 \hline
 $0.012$ & $2.12\!\pm\!0.07$ & $0.94\!\pm\!0.03$ & $1.09\!\pm\!0.03$ & $1.63\!\pm\!0.04$ \\
 $0.008$ & $2.22\!\pm\!0.11$ & $0.90\!\pm\!0.05$ & $1.06\!\pm\!0.04$ & $1.59\!\pm\!0.06$ \\
 $0.004$ & $2.33\!\pm\!0.23$ & $0.86\!\pm\!0.09$ & $1.13\!\pm\!0.06$ & $1.52\!\pm\!0.09$ \\
 $0$      & $2.47\!\pm\!0.35$ & $0.81\!\pm\!0.12$ & $0.95\!\pm\!0.07$ & $1.43\!\pm\!0.11$ \\
 $-0.004$  & $2.30\!\pm\!0.42$ & $0.87\!\pm\!0.16$ & $0.84\!\pm\!0.06$ & $1.26\!\pm\!0.09$ \\
 $-0.008$  & $2.41\!\pm\!0.53$ & $0.83\!\pm\!0.18$ & $0.79\!\pm\!0.06$ & $1.19\!\pm\!0.09$ \\
 \hline
 \end{tabular}\label{table-1}} 
\end{table}
We find that $D_f$ at $0.008\!\leq\!{\it \Delta}p\!\leq\!0.012$ is  $D_f\!=\!2.1\!\sim 2.4$.  This value is consistent with the experimentally obtained number with partially polymerized membranes \cite{Chaieb-etal-2006PRL,Chaieb-Malkova-Lal-2008JTB}. In the collapsed phase at ${\it \Delta}p\!=\!0.012$, our analysis gives $D_f\!=\!2.12\!\pm\!0.07$.  This number is larger than $D_f\!=\!2$ only slightly. This is due to incomplete collapse which is observed in our model at ${\it \Delta}p\!=\!0$ . 

A previous study using the non-perturbative renormalization group formalization predicts that $\nu_{c}\!=\!0.8$ and  $\nu_{ct}^G\!=\!0.78$, corresponding to the radius $R_c$ and the tubule thickness $R_G$ \cite{Radzihovsky-SMMS2004}. These values may not be directly compared to our obtained numbers, since our model includes an isotropic bending rigidity while the model in Ref. \refcite{Radzihovsky-SMMS2004} assumes an anisotropic bending rigidity.  However, the result $\nu_{{\rm x}^2}\!=\!0.81(12)$ at ${\it \Delta}p\!=\!0$ is in a reasonable agreement with the previously obtained theoretical predictions.   

The exponent $\nu_{{\rm R}^2}$ can be compared with $\bar{\nu}_{\rm v}$ and $\nu_{\rm v}$, which are defined by 
\begin{equation}
\label{V_scale}
 V\sim N^{\frac{3}{2}\bar{\nu}_{\rm v}}, \quad V\sim N^{\nu_{\rm v}} \quad (N\to \infty).
\end{equation}
It is expected that $\nu_{{\rm R}^2}\!=\! \bar{\nu}_{\rm v}$ in the inflated phase of the fluid vesicle model \cite{Gompper-Kroll-PRA1992,Gompper-Kroll-EPL1992}, and the value of $\nu_{{\rm R}^2}(=\! \bar{\nu}_{\rm v})$ corresponds to $2\nu_+$ in Ref. \refcite{Gompper-Kroll-PRA1992,Gompper-Kroll-EPL1992}. The obtained exponents $\bar{\nu}_{\rm v}$ and  $\nu_{\rm v}$ holds the relation $\nu_{{\rm R}^2}\!=\! \bar{\nu}_{\rm v}$  in the tubular phase close to the CT transition point. Indeed, we find that $\nu_{{\rm R}^2}\!=\! \nu_{\rm v}(=\!0.8\!\sim\!0.85)$ in the region $-0.008\!\leq\!{\it \Delta}p\!<\!0$ and its value is clearly smaller than $1$ (Table \ref{table-1}). This is consistent with our observation that the tubular surfaces are different from the branched polymer surfaces of the fluid vesicle model, where $\nu_{{\rm R}^2}\!=\! \nu_{\rm v}\!=\!1$ is satisfied \cite{Gompper-Kroll-PRA1992,Gompper-Kroll-EPL1992}.  
 In the crumpled phase at ${\it \Delta}p\!\geq\!0.004$, we clearly see  $\nu_{{\rm R}^2}\!\not=\! \bar{\nu}_{\rm v}$. We also find 
\begin{equation}
\label{expected-scaling}
\bar{\nu}_{\rm v}=0.95\pm0.07
\end{equation}
at $ {\it \Delta}p=0$, which is close to the CT transition point as mentioned above. 
This value is identical to $\nu\!=\!0.95(5)$ in Ref. \refcite{BOWICK-TRAVESSET-EPJE2001}. This implies that the surface at the CT transition of the model in this paper is in the same phase as the flat phase of the model in Ref.
\refcite{BOWICK-TRAVESSET-EPJE2001}.  However this value in Eq. (\ref{expected-scaling})  is relatively smaller than $2\nu_+\!\simeq\!1.16$  \cite{Gompper-Kroll-PRA1992,Gompper-Kroll-EPL1992} and $2\nu_+\!\simeq\!1.12$ \cite{Dammann-etal-JPIF1994}. 
 
\section{Summary and Conclusion}\label{Conclusion}
We have numerically studied a self-avoiding meshwork model on lattices that consist of connection-fixed triangles. The model has nonzero in-plane shear rigidity at each triangle like the connection-fixed model with the Gaussian bond potential. We have found that the model undergoes a second order transition between the crumpled and tubular phases (CT transition) at ${\it \Delta}p\!\simeq\!0$ under a constant $\kappa$. The parameter $\kappa$ is fixed at $\kappa\!=\!0.64$ in order for the CT transition to occur at ${\it \Delta}p\!\simeq\!0$. Also we have found that the surface at the CT transition is relatively inflated, and this observation is confirmed by the size exponents. 

\section*{Acknowledgments}
This work is supported in part by Promotion of Joint Research, Nagaoka University of Technology. The autor H.K. would like to thank J.-P. Kownacki and D. Mouhanna for reminding him the CT transition, and he is also grateful to Koichi Takimoto in Nagaoka University of Technology for careful reading of the manuscript. We thank Hiroki Mizuno for the support of computer analyses. We acknowledge Takashi Matsuhisa for comments.




\end{document}